\documentclass[a4paper]{jpconf}
\usepackage{graphicx}
\usepackage{amsmath}
\usepackage{amssymb}
 \usepackage{url} 
 \newcommand{\open}{\sphericalangle}
\begin{document}
\title{Phenomenology of Dihadron Fragmentation Function.}

\author{A.~Courtoy}

\address{CONACyT---Departamento de F\'isica, Centro de 
Investigaci\'on y de Estudios Avanzados, Apartado Postal 14-740, 07000 
Ciudad de M\'exico, M\'exico}

\ead{acourtoy@fis.cinvestav.mx}

\begin{abstract}
We report on the phenomenological results obtained through Dihadron Fragmentation Functions related processes. 
In 2015, an update on the fitting techniques for  the Dihadron Fragmentation Functions  has led to an improved extraction of the transversity PDF and, as a consequence, the nucleon tensor charge. We discuss the impact of the determination of the latter on search for physics Beyond the Standard Model, focusing on the error treatment. We also comment on the future of the extraction of the subleading-twist PDF $e(x)$ from JLab soon-to-be-released Beam Spin Asymmetry data.
\end{abstract}

\section{Introduction}

\vspace{.2cm}

Dihadron Fragmentation Functions (DiFF) encode information about the fragmentation process of a quark into a hadron pair (plus something else, undetected). Fragmentation has been hardly studied through models for pion pairs, the main knowledge about those DiFFs  being for now fits from data.

The rich phenomenology associated to dihadron fragmentation compares to the well-known Semi-Inclusive Deep Inelastic Scattering (SIDIS), in which a single hadron is produced in the current fragmentation region. In these proceedings, we will review the expectations rather than the acknowledged successes of the subset of Dihadron data, that can be found elsewhere~\cite{Radici:2015mwa,Bacchetta:2012ty}. These proceedings will focus on the error treatment.

\section{The future of Dihadron FF from data}

\vspace{.2cm}

The unpolarized DiFF have been extracted by fitting the pair distribution simulated by a Monte Carlo event generator~\cite{Courtoy:2012ry,Radici:2015mwa}, to compensate for the absence of data on multiplicities. The resulting error on the parametrization of the unpolarized DiFF is obviously small. On the other hand, once the unpolarized DiFF parameterized, the chiral-odd DiFF can be extracted from the Artru--Collins asymmetry at Belle. The factorization of the process
$e^+ e^- \to (\pi^+ \pi^-)_{\mbox{\tiny jet}} (\pi^+ \pi^-)_{\overline{\mbox{\tiny jet}}} X$ is valid in the kinematical regime $P_h^2 = M_h^2 \ll Q^2=-q^2 \geq 0$ and $q = k - k'$ the space-like momentum transferred, with the total momentum $P_h = P_1 + P_2$.

\begin{figure}
\centering
\includegraphics[width=8cm]{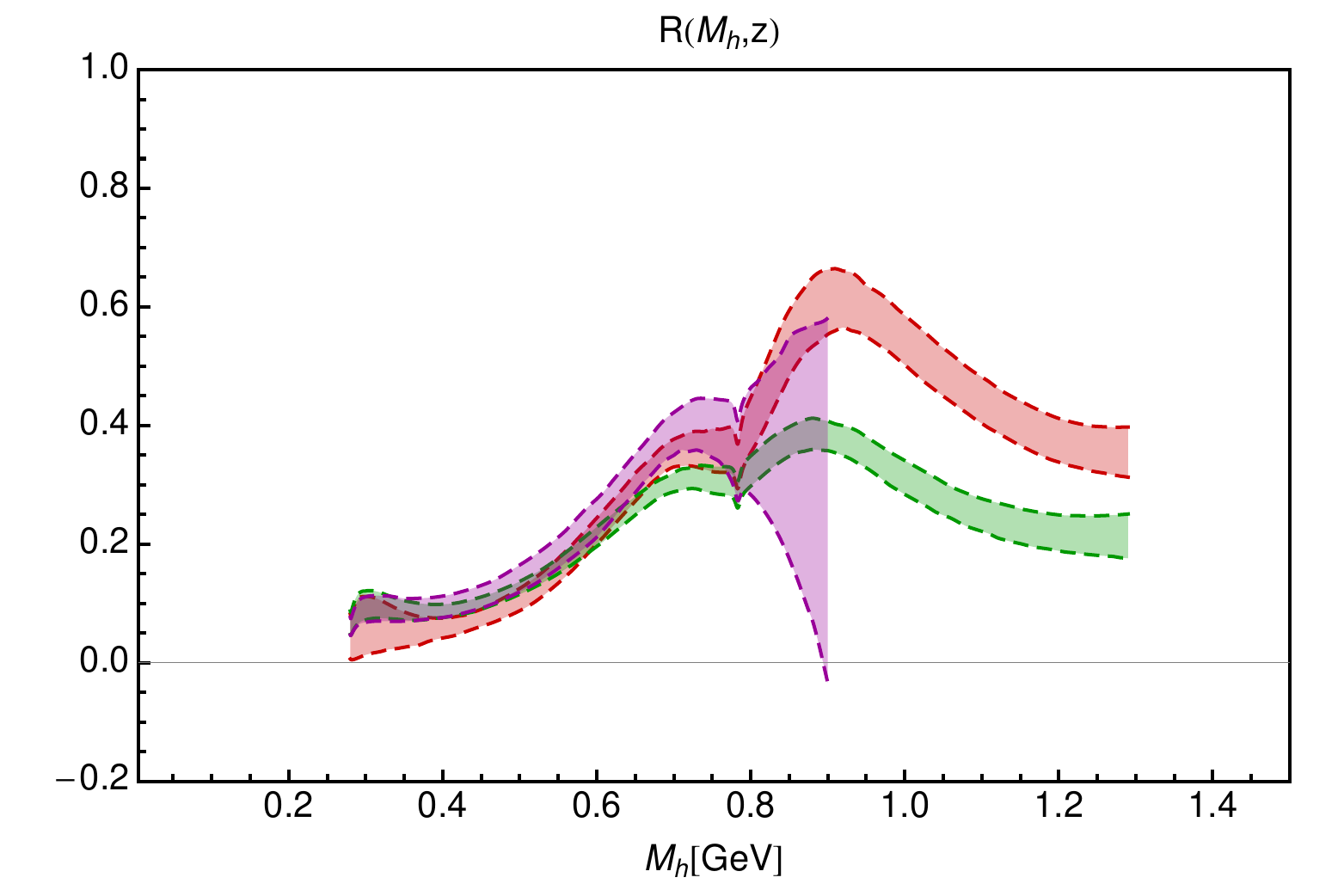} 
\caption{The ratio $R(z, M_h)$ as a function of $M_h$ at $Q_0^2=1$ GeV$^2$ for three different $z=0.25$ (shortest band), $z=0.45$ (lower band at $M_h \sim 1.2$ GeV), and $z=0.65$ (upper band at $M_h \sim 1.2$ GeV), with the value $\alpha_s (M_Z^2) = 0.139$ used in the QCD evolution equations. } 
\label{fig:H1Mh}
\end{figure}

 Belle data for the Artru--Collins asymmetry led to the extraction ---and, hence, parameterization--- of the leading-twist DiFFs~\cite{Courtoy:2012ry,Radici:2015mwa}, illustrated in Fig.~\ref{fig:H1Mh} through their ratio
\begin{equation}
R(z, M_h) = \frac{|\bf{R}|}{M_h}\, \frac{H_1^{\open\, u} (z, M_h; Q_0^2)}{D_1^u (z, M_h; Q_0^2)} \; ,
\label{e:R}
\end{equation}
 the relevant variables being the pion pair invariant mass,  $M_h$, and the sum of the fractional energies carried by the two final hadrons, $z$.
 
 While the first fitting approach was based on the usual Hessian statistics with a variation of the chisquare equal to unity~\cite{Courtoy:2012ry}, the updated version incorporates a Neural Network preparation, {\it i.e.} the normal generation of data replicas within the ---$1\sigma$---  experimental errors~\cite{Radici:2015mwa}. This second technique insures the proper treatment of errors outside the data range, thus avoiding the vanishing of the errors at the endpoints, where they should be maximal. We have noticed that both fitting techniques, still based on adhoc functional forms, agree with each other as far as the chiral-odd DiFF is concerned.
 \\
 
 In principle, the unpolarized $D_1$ should be extracted by global fits of the unpolarized cross section, in the same way as it is done for single-hadron fragmentation~\cite{deFlorian:2007aj}. The  first analysis of di-hadron multiplicities is ongoing at COMPASS~\cite{Makke:2013dya} and an analysis of Dihadron SIDIS multiplicities has been proposed at CLAS12~\cite{prop_ht_silvia}.
 The expression for the multiplicities can be written in  terms of PDFs and DiFFs as
\begin{eqnarray}
M^{h}(z, M_{h}, x; Q^2)&=&
\frac{\sum_q \, e_q^2\, f_1^q(x; Q^2)\, D_1^q(z, M_{h}; Q^2)}{\sum_q \, e_q^2\, f_1^q(x; Q^2)}\quad ,
\end{eqnarray}
where $f_1$ is the well-known collinear unpolarized PDF and $D_1$ is the unpolarized DiFF. It corresponds to the ratio of 
the number $N^{\pi^+\pi^-}$ of produced pion pairs in SIDIS and the number of DIS events $N^{\mbox{\tiny DIS}}$ for the same kinematics. 
With the future CLAS12 data, we will analyze the 4-dimensional binning spanning a wide kinematical range in $(z, M_h, x, Q^2)$, for both proton and deuteron targets, allowing for a flavor separation of the valence distributions. 
The development expected from such a measurement and statistical analysis will qualitatively improve the present estimates for a number of processes. 
As most of the Distribution/Fragmentation Functions, DiFF are believed to be universal. This characteristic has recently been studied and reinforced through a confirmation on the role of transversity in proton-proton collisions~\cite{Radici:2016lam}.

\section{Completing the collinear twist-2 picture}

\vspace{.2cm}

Combining the parametrization obtained for DiFFs to HERMES and COMPASS data on dihadron SIDIS allowed for the extraction of the last leading-twist PDF, {\it i.e. the transversity PDF}~\cite{Radici:2015mwa,Bacchetta:2012ty}. While the fit of the transversity is an achievement by itself, though affected of a large uncertainty outside the data kinematical range, its first Mellin moment is also of great interest. The tensor charge\footnote{Note that the word charge here is a misuse of language.} is obtained by integrating the transversity PDF over the physical support in $x$,
\begin{eqnarray}
\delta q_v (Q^2) &= &\int_0^1 dx \, h_1^{q_v} (x, Q^2) \; . 
\label{e:tensch}
\end{eqnarray}
The error we mentioned above, consequence of the extrapolation of the PDF outside the data range, is the main source of uncertainty on the determination of the tensor charge, as shown in Table 3 of Ref.~\cite{Radici:2015mwa}. Another source of error consists in the choice of the functional form for the transversity PDF. This problem has been taken care of, in a first step, considering three different functional forms,  each with a growing number of free parameters. The value for the isovector tensor charge $g_T = \delta u_v - \delta d_v$   for the functional form related to the so-called {\it flexible scenario} with $\alpha_s (M_Z^2)=0.125$ is, at $1\sigma$,
\begin{eqnarray}
g_T = 0.81 \pm 0.44\quad \mbox{ at} \quad Q^2=4 \, \mbox{GeV}^2\quad.
\label{eq:gtpavia}
\end{eqnarray}
 It is in agreement with lattice determinations as well as with the other extractions from hadronic phenomenology \cite{Anselmino:2013vqa,Goldstein:2014aja,Kang:2015msa}, though the absolute value is slightly smaller than the lattice's. Due to its non-perturbative nature, the nucleon structure can only be unveiled using complementary methods such as effecive field theories, lattice calculations, models for the nucleon structure, Schwinger-Dyson based techniques and phenomenological extraction of observables from data. As such, the lattice field theory and the data based determinations play a similar, yet uncorrelated, role. 
\\

%
%


The role of the tensor aspect of the nucleon becomes clear in beta decay where the hadronic characterization of the process is encoded through, among others, the tensor form factor ---which value at zero momentum transfer is nothing else than the tensor charge. While there is no leptonic counterpart of the tensor structure in the Standard Model, all quark bilinear Lorentz structures can be introduced through
an effective Lagrangian which is relevant for beta decay
observables. The beta decay observables, {\it e.g.} $b_0^+$ and the Fierz term $b$, involve products of the Beyond the Standard Model (BSM) couplings, $\epsilon_i$, and
the corresponding hadronic charges, $g_i$.
The scalar (S) and tensor (T)
operators, in particular, contribute linearly to the beta decay
parameters through their interference with the SM amplitude,
and they are, therefore, more easily detectable. 
We focus here on the following low-energy effective  interaction (see Ref.~\cite{Cirigliano:2013xha} for a review)
\begin{eqnarray}
\Delta{\cal L}_{\rm eff} 
&=&
- C_S \bar{p}n\, \bar{e}(\mathbb{I}-\gamma_5)\nu_e
- C_T \bar{p}\sigma_{\mu\nu}n\, \bar{e}\sigma^{\mu\nu}(\mathbb{I}-\gamma_5)\nu_e\quad,
\label{eq:leffq2} 
\end{eqnarray}
where $C_S =   G_F V_{ud} \sqrt{2} \epsilon_{S}  g_{S}$, and $C_T =   4 G_F V_{ud} \sqrt{2} \epsilon_{T}  g_{T}$ and $G_F\equiv \sqrt{2}g^2/(8 M_W^2)$ is the tree-level definition of the Fermi constant.
In  the SM, the $\epsilon_S$ and $\epsilon_T$ coefficients 
vanish leaving the well-known $(V-A)\times(V-A)$ structure generated by the exchange of a $W$ boson. 

Precise measurements in beta decay set strong bounds on the combination $g_T\epsilon_T$ obtained through global fits~\cite{Pattie:2013gka,Wauters:2013loa},
\begin{eqnarray}
\left | g_T\epsilon_T \right | < 6\cdot 10^{-4}~~~~~~\mbox{(90\% C.L.)}~,
\label{eq:gtet}
\end{eqnarray}
which is expected to be improved by the next generation of experiments. 

In order to extract a bound on the Wilson coefficient $\epsilon_T$ from Eq.~(\ref{eq:gtet}) it is necessary to know the value of the tensor charge $g_T$. 
Hence, the sensitivity of beta decay measurements to an exotic tensor interaction depends on our knowledge of the tensor charge. 
The impact of data based determinations of the tensor charge has been studied in Ref.~\cite{Courtoy:2015haa}. Given the  sensitivity bounds on the Fierz term $b$, an error of about $10-15\%$ on the tensor charge is estimated to be sufficient. The values for $g_T$ mentioned above are still far from that precision. 

Originally, in Ref.~\cite{Courtoy:2015haa}, an alternative
to the standard Hessian evaluation was considered. The argument 
that 
both the lattice QCD and experimental extractions of the couplings from $\beta$-decay
are
affected by systematic or theoretical uncertainty invalidates the assumption of a gaussian distribution of the error around the central value. 
In order to deal with this situation we followed Ref.~\cite{Bhattacharya:2011qm} and we calculated the confidence interval on $\epsilon_T$ using the so-called R-Fit method~\cite{Hocker:2001xe}. In this scheme the theoretical likelihoods do not contribute to the $\chi^2$ of the fit and the corresponding QCD parameters take values within certain \lq\lq allowed ranges". In our case, this means that $g_T$ is restricted to remain inside a given interval, e.g. $0.37 \leq  g_T  \leq 1.25$  for the current determination from di-hadron SIDIS. 
The chisquare function is then given by
\begin{equation}
\chi^2 (\epsilon_T)  =  {\rm min}_{g_T} \left( \frac{\left[ g_T \epsilon_T \right]^{\rm exp}    - g_T \epsilon_T}{\delta \left[ g_T \epsilon_T \right]^{\rm exp}}\right)^2\quad,
\label{eq:chirfit}
\end{equation}
where the minimization is performed varying $g_T$ within its allowed range. In this approach, the bound on $\epsilon_T$ depends only on the lower limit of the tensor charge, as long as the experimental determination of $g_T\epsilon_T$ is compatible with zero at $1\sigma$.  In particular, the tensor charge given by Eq.~(\ref{eq:gtpavia})  leads to a bound $|\epsilon_T|<0.00162$. This bound is the largest found through hadron phenomenology (see Fig.~2 of Ref.~\cite{Courtoy:2015haa}) but the error on $g_T$ is expected to decrease of about $10\%$ with the new JLab data.
For comparison, the bound obtained from the analysis of LHC data carried out in Ref.~\cite{Gonzalez-Alonso:2013uqa} is $|\epsilon_T|<0.0013$.

This error analysis presents shortcomings, {\it e.g.} the fact that Eq.~(\ref{eq:chirfit}) only depends on the lower bound on the tensor charge. The principal criticism would be that the R-fit method is a global method for data fitting, used here for the determination of a single parameter based on a fit output, Eq.~(\ref{eq:gtet}).

Another alternative analysis could be implemented taking advantage on the replica technique of the tensor charge determination in DiFF related processes. The bound given by Eq.~(\ref{eq:gtpavia}) is nothing else then the 64 most central values, output of the 100 fitted data replicas. The $90\%$ C.L. is easily obtained by discarding only ten outer values (5 upper and 5 lower). Generating normally 100 replicated values for the product $\epsilon_T \, g_T$ around zero with a variance of $3\times 10^{-4}$, we obtain a Monte Carlo-like estimate of the error, fully at $90\%$ C.L.: $|\epsilon_T|<0.00139$.
\\

\begin{figure}
\center
\includegraphics[width=8.cm]{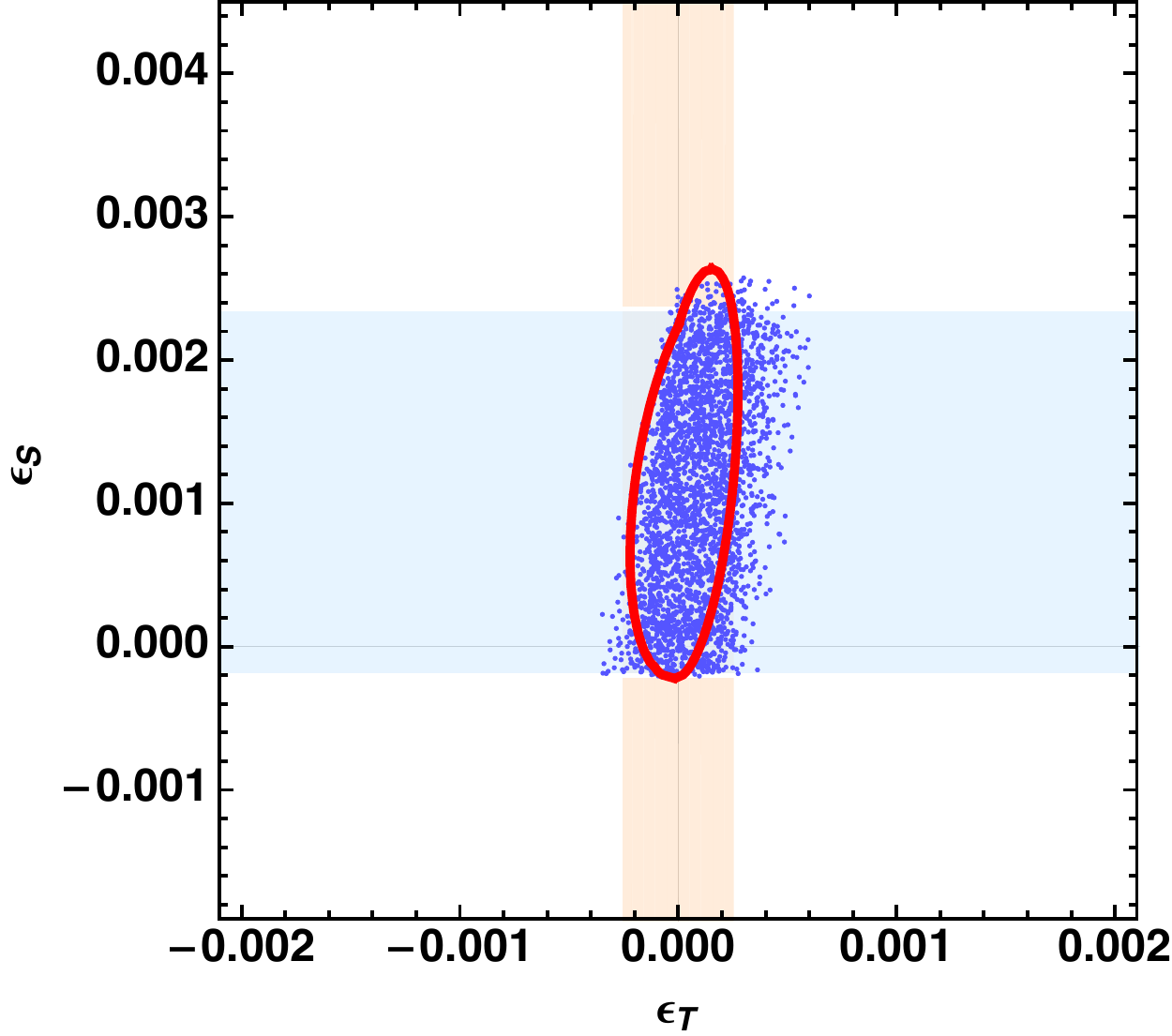}
\caption{The $(\epsilon_T, \epsilon_S)$ plane. Allowed values at $1\sigma$ evaluated in three different statistical methods. Scatter plot (blue dots),  contour plot (red curve) and Hessian method for $\epsilon_T$ and $\epsilon_S$ in orange and light blue, respectively. }
\label{fig:et}
\end{figure}

Extending the previous analysis to two dimensions, we can use the bound on the relevant observables obtained from the decay rate measurement, {\it e.g.}, $b_0^+$ from super-allowed Fermi transition as well as the Fierz term $b$. The latter can be expressed in terms of the factorized coupling~\cite{Bhattacharya:2011qm}
\begin{eqnarray}
\label{b:eq}
b& = &  \frac{2}{1+ 3\lambda^2} \left[ g_S \epsilon_S  - 12 g_T \epsilon_T \lambda \right]  < 10^{-3}\nonumber\quad,\\
b_0^+&=& -2 g_S \epsilon_S < 0.0022(26)\quad,
\label{eqs:bb0}
\end{eqnarray}
with $\lambda = g_A/g_V$. The bound on $b_0^+$ is given in Ref.~\cite{Hardy:2008gy} and the precision limit on the Fierz term corresponds to future experiments expectations. 

A simple excercise to compare various techniques for the error treatment consists in calculating the propagation of error for the $(\epsilon_T, \epsilon_S)$ plane based on Eqs.~(\ref{eqs:bb0}). We propose a simplified version\footnote{Simplified as compared to accurate global fit analyses.} for the sake of illustration, in which we also assume that there is no error on $\lambda$.
In Fig.~\ref{fig:et}, we show the allowed region for the plane $(\epsilon_T, \epsilon_S)$ using three different statistical techniques, namely the Hessian method, the R-fit method and the space scan of the permitted values of parameters.  We have used   $g_S=1.02\pm 0.11$ for the scalar charge~\cite{Gonzalez-Alonso:2013ura}.

The first technique is the most commonly used error propagation. The error on each effective coupling is given by 
\begin{eqnarray}
\Delta^2 \epsilon_S&=&\frac{1}{4}\left [ \left( \frac{\Delta b_0^+}{g_S}\right)^2+  \left( \frac{\Delta g_S}{g_S^2}\right)^2 \, b_0^{+\,2}\right ] \quad, \nonumber\\
\Delta^2 \epsilon_T&=&\frac{1}{(12\, \lambda)^2} \left[\Delta^2 b  \left(\frac{ 1+3\lambda^2}{2\,g_T}\right)^2 +\left( \frac{\Delta b_0^+}{2\,g_T}\right)^2
+\left(\frac{\Delta g_T}{ g_T^2}\right)^2 \left[\left( \frac{b(1+3\lambda^2)}{2}\right)^2+\left(\frac{b_0^+}{2}\right)^2\right]\right]\quad.\nonumber\\
\end{eqnarray}
Notice that, here, $b=0$. The result is given by the light orange and blue bands in Fig.~\ref{fig:et}. There are no correlation taken into account in this technique.

The second approach consists in using the  R-Fit method again. In this scheme,  $g_{S/T}$ is restricted to remain inside the  interval defined by its error at 1$\sigma$. Notice that all values inside this range are treated on an equal footing.
The chisquare function is then given by
\begin{eqnarray}
&&\chi^2 (\epsilon_S, \epsilon_T)  \nonumber\\
&&=  {\rm min}_{g_{S,T}}\left[ \left( \frac{\left(b_0^+\right)^{\rm exp}    - \left(-2\, g_S\epsilon_S\right)^{\rm theo} }{\left(\Delta b_0^+\right )^{\rm exp}}\right)^2~+\left( \frac{b^{\rm exp}    - 2/(1+ 3\lambda^2)\times \left[ g_S \epsilon_S  - 12 g_T \epsilon_T \lambda \right] ^{\rm theo}}{\left(\Delta b\right )^{\rm exp}}\right)^2 \right]\quad.\nonumber\\
\label{eq:rfitchi2}
\end{eqnarray}
Again, $b^{\rm exp}=0$. The $1\sigma$ contour is found equating  Eq.~(\ref{eq:rfitchi2}) to $1$.  It is depicted by the red contour in Fig.~\ref{fig:et}.

Finally, we show a scan of the allowed $(\epsilon_S, \epsilon_T)$ plane solving the equations for $b$ and $b_0^+$ varying normally $g_{S,T}$ at $1\sigma$.  It corresponds to the area of blue points in Fig.~\ref{fig:et}. The three techniques seem to be compatible for the present case, with the available precision.

\section{Twist-3 from DiFF}

\vspace{.2cm}

In the previous Section, we have presented an analysis based on the extracted tensor charge~\cite{Radici:2015mwa} together with the scalar charge from isospin breaking analysis~\cite{Gonzalez-Alonso:2013ura}. Ideally, we would like to determine the scalar charge from phenomenology as well. 

Though in a more elusive way, the scalar charge is related to a twist-3 PDF, $e(x)$.  QCD equations of motion allow to decompose the chiral-odd twist-3 distributions into three terms%
\begin{eqnarray}
e^q(x)&=&e_{\mbox{\tiny{loc}}}^q(x)+e_{\mbox{\tiny{tw-3}}}^q(x)+e_{\mbox{\tiny{mass}}}^q(x)\quad.
\end{eqnarray}
The first term comes from the local operator and is exactly the contribution related to scalar charge,
\begin{eqnarray}
e_{\mbox{\tiny{loc}}}^q(x)&=&\frac{1}{2M}\, \int \frac{d\lambda}{2\pi}\, e^{i\lambda x}\langle P| \bar{\psi}_q(0)\psi_q(0)|P\rangle\quad,\nonumber\\
&=&\frac{\delta(x)}{2M}\langle P| \bar{\psi}_q(0)\psi_q(0)|P\rangle\quad;
\end{eqnarray}
the second term is a genuine twist-3 contribution, {\it i.e.} pure quark-gluon interaction term ; while the last term is related to the current quark mass.

The sum rule related the chiral-odd twist-3 PDF to the scalar charge is similar to Eq.~(\ref{e:tensch}). Only the {\it local} term contributes to the first Mellin moment, {\it i.e.} the scalar charge is nothing else than $e(x=0)$.  That delta-function singularity follows from chiral symmetry and the existence of non-vanishing quark condensate~\cite{Wakamatsu:2003uu}.

\begin{figure}
\centering
\includegraphics[width=8cm]{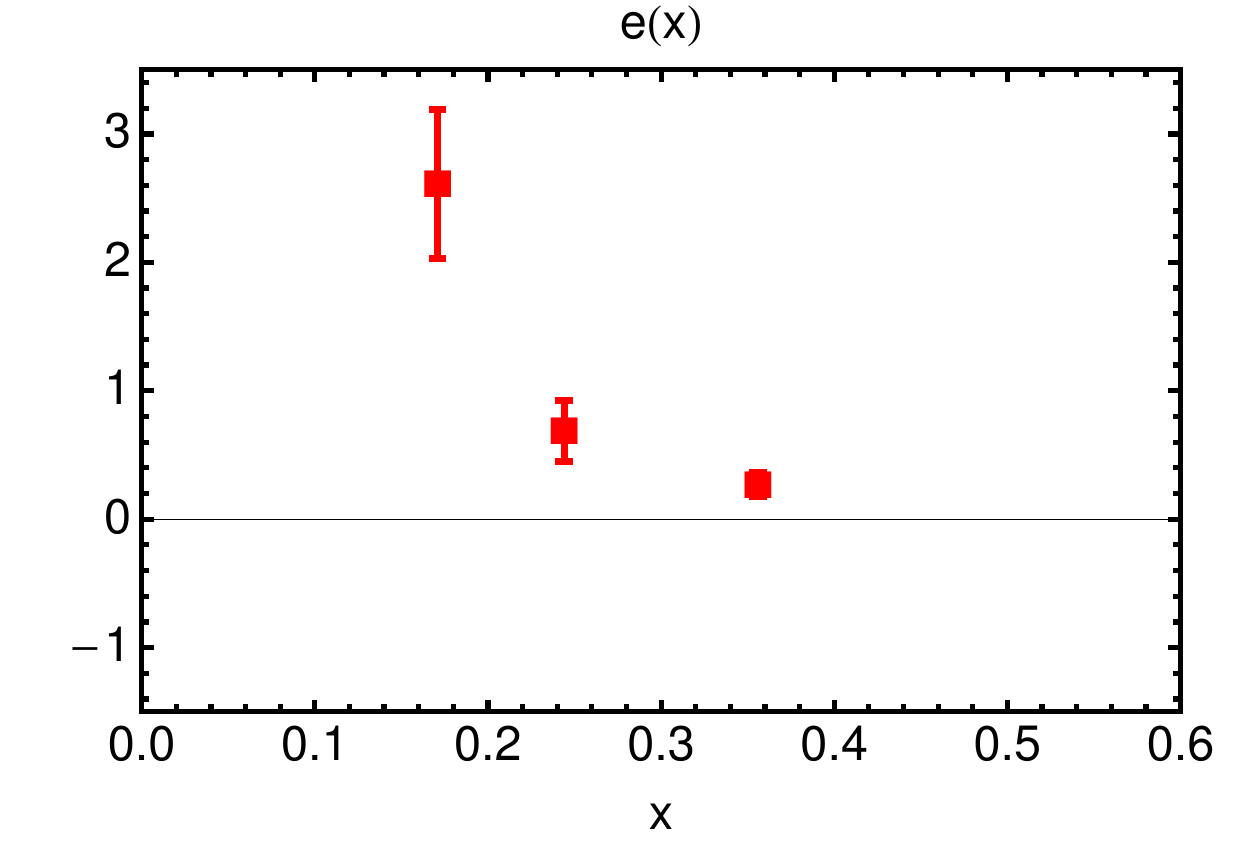} 
\caption{ The valence combination $e(x)\equiv4 e^{u_V}(x_i, Q_i^2)/9-e^{d_V}(x_i, Q_i^2)/9$ (see text). The error bars correspond to the propagation of the experimental and DiFF errors. } 
\label{fig:eww}
\end{figure}

Being a subleading contribution, this twist-3 PDF is hardly known. It is however accessible in single-~\cite{Efremov:2002ut} and di-hadron SIDIS.  In the latter case, the chiral-odd partner of $e(x)$ is the chiral-odd dihadron fragmentation function, {\it i.e.} $H_1^{\sphericalangle}$. 
An extraction of the twist-3 PDF, $e(x)$, through the  analysis of the  preliminary data~\cite{Pisano:2014ila} for the $\sin\phi_R$-moment of the beam-spin asymmetry for  dihadron Semi-Inclusive DIS at CLAS at 6 GeV was proposed in Ref.~\cite{Courtoy:2014ixa}. It is illustrated in Fig.~\ref{fig:eww} in a specific scenario ---in which the asymmetry is dominated by the term containing the twist-3 PDF.  By relaxing that hypothesis, we can define other scenarios: the function $e(x)$  always  results non-zero. 

In the next few years, future data at CLAS12~\cite{prop_ht_silvia}   will be analyzed that bring us more statistics and information on that particular function. The determination of the scalar charge will be a delicate task as the value of the function at $x=0$ is experimentally unaccessible. The solution will be to appeal to other properties of the subleading-twist function together with advanced fitting techniques.

The determination of the scalar charge represents a challenge for the hadronic physics community but an opportunity to experimentally explore the consequence of chiral symmetry on the one hand. On the other hand, the search for a new fundamental scalar interaction, Beyond the Standard Model, would benefit from the outcome of this line of research, complementing the results obtained by the lattice community.

\section{Conclusions}

\vspace{.2cm}

The phenomenology of Dihadron Fragmentation Functions integrates the idea that more structures and information become accessible when incorporating a dependence on the transverse momentum. The elegance of DiFF results in  less intricated and more easily taken care of observables, namely collinear objects and simple products, as opposed to convolutions. This particular feature allows for more flexibility in their statistical analysis. The parameterization of the third leading-twist PDF, {\it the transversity}, marked the first stage of an alternative approach to transverse momentum dependent functions. Then followed the tensor charge determination, made possible thanks to the flavor structure of the analyzed observables. The role of dihadron in proton-proton collision has been studied as well. Subleading processes involving DiFF are being analyzed and will be extended to JLab@12. 

Twist-3 functions encode invaluable information on the low energy dynamics of quarks and gluons, so far hardly explored. The nonperturbative nature of nucleons influences numerous observables that involve them. Chiral symmetry, the existence of a delta-function singularity, the pion-nucleon sigma-term, all pieces of a same puzzle that could further contribute to searches of New Physics, {\it e.g.} \cite{Ellis:2008hf}.
The possibility of obtaining the scalar and tensor charges directly from experiment with sufficient precision, 
not having to deal with purely theoretical uncertainties of these quantities,
 gives a different weight to searches for BSM involving nucleons. 
 
 The results presented here concerning the determination of the tensor charge through Dihadron Fragmentation Functions are complemented by the Deeply Virtual Meson Production analysis~\cite{Goldstein:2014aja} and the Single-hadron SIDIS approach~\cite{Anselmino:2013vqa}.

To conclude, we would like to restate that dihadron SIDIS offers a unique way to access collinear functions,  subleading functions being of  particular interest. Though there are few data now,
future analyses  ---{\it e.g.} dihadron multiplicities at COMPASS--- and experiments ---especially in CLAS12 and SoLID at JLab--- will increase the data set, allowing for an improved knowledge on both DiFFs at leading and subleading-twist and collinear leading and subleading PDFs. 

\ack 
The author acknowledges her  co-authors from Pavia, A. Bacchetta, M.~Guagnelli and M.~Radici, discussions and broad collaboration with S.~Liuti and S.~Pisano.  A.C. would like to thank the collaborations with SoLID and CLAS12 in general. This work is supported by a C\'atedra CONACyT contract.

\end{document}